\documentclass[aps,prl,twocolumn,groupedaddress]{revtex4-1}

\usepackage{graphicx}
\usepackage{subcaption}
\usepackage{float}
\begin{document}


\title{Laser Pulse-Electron Beam Synergy: An Approach to Higher Electron Energy Gain in Laser Wakefield Accelerators}


\author{S. Barzegar and A. R. Niknam}
\email[]{a-niknam@sbu.ac.ir}
\affiliation{Laser and Plasma Research Institute, Shahid Beheshti University, Tehran 1983969411, Iran}


\date{\today}

\begin{abstract}
A new scheme for injection and acceleration of electrons in wakefield accelerators  is suggested based on the co-action of a laser pulse and an electron beam. This synergy leads to stronger wakefield generation and higher energy gain in the bubble regime. The strong deformation of the whole bubble leads to electron self-injection at lower laser powers and lower plasma densities. To predict the practical ranges of electron beam and laser pulse parameters an interpretive model is proposed. The effects of altering the initial electron beam position on self-trapping of plasma electrons are studied. It is observed, a high quality (25 fs), high charge (340 pC ), 1 GeV electron  bunch is produced by injection of a 280 pC electron beam in the decelerating phase of the 48 TW laser driven wakefield.  \end{abstract}


\maketitle

Acceleration of electrons to relativistic energies in the blowout regime of laser wakefield accelerators in very short distances is one of the most promising schemes \cite{pukhov,geddes,Thomas,maksimchuk,Ralph,rezaei}.
In this technique, the ponderomotive force of laser pulse forms extremely nonlinear wakefields by expelling the whole background plasma electrons and generating a near-spherical cavitated region that supports perfect accelerating and focusing fields \cite{rosenzweig}. Providing a path to produce high-quality, high energy  electron bunches using different methods of injecting electrons in plasma wakefield \cite{suk,pak,lehe,oz} and  accelerating them to GeV range \cite{tsung,pollock,jin} has drawn wide attention recently.  

The basic requirements to achieve high-quality multi-GeV electron bunches are producing high gradient accelerating field structure and the way of injecting electrons into them. As we know, lower plasma densities enhance the maximum energy gain of electrons  because at lower densities the phase velocity of the wake increases, thereby increases the dephasing length \cite{schroeder}. However, it is necessary to use higher laser pulse energy and longer accelerator lengths in this regime and these conditions make the guiding of laser pulse more challenging \cite{gonsalves}. For self-guided propagation the laser  power must be higher than a critical power, $P\succsim P_c$ \cite{sun}, where $P_c=17{\omega_{0}}^2/{\omega_{p}}^2 [GW] $, $\omega_{0}$ and $\omega_{p}$ are laser frequency and plasma frequency, respectively. This  indicates that as the density decreases we need to increase the laser intensity to maintain self-guiding because the technology for making the meter-scale plasma channels for external guiding is not developed yet. Although, as described in Refs. \cite{decker,vieira,lu3} the leading edge of the laser pulse due to the refractive index modification i.e., combined effects of electron density decrease and relativistic mass increase, locally pump depletes and the rate of etching will be faster for higher intensities. Eventually, the more the laser pulse energy is the faster will lose its energy and can not guide in plasma and excite a stable wake for a long distance. On the other hand, laser pulse intensity cannot be too large if one needs high efficiency \cite{lu3}. Therefore, using moderately large intensity, short-pulse laser produces more stable accelerating structure.  However, wakefield strength depends on the input laser power. Therefore,  discovering new methods for generating stronger wakefield structures and subsequently extending the electron beam energy gain for moderate laser intensities seems an essential subject in the development of plasma-based accelerators.

This paper considers the stable guiding of an intense laser pulse which is followed by a short electron bunch. An electron bunch is externally injected into the decelerating field of plasma bubble and severely modifies the structure of plasma wakefield. Injecting electron beams accelerated in a first stage to a second wakefield stage that is driven by a different laser pulse is investigated experimentally in multi-stage accelerators before \cite{pollock,luo}. We indicate here that in this new regime, the bubble as a whole is strongly affected. The modification of the wake strengthen laser wakefield accelerating phase ($\sim 180$ GeV/m) and considerable improvement in energy gain ($\sim 250$ MeV)  can be achieved compared to laser wakefield accelerators (LWFAs) and plasma wakefield accelerators (PWFAs).

On the other side, self-trapping of plasma electrons at low electron densities and accelerating them to GeV range for presently available laser systems (100  TW) look likes difficult. Therefore, generating quasi-mono-energetic beams of electrons from plasma are considered as an important concern. We also report a new scheme for plasma electron self-trapping into plasma wakefield using hybrid accelerator which happens at lower laser powers and densities than would be possible. Phase slippage (dephasing) between the decelerating electric field  of bubble and injected electron (electron beam velocity is higher than wake velocity in a plasma), leads to bubble evolves constantly and the expansion of the bubble triggers injection \cite{kalmykov,yi,kostyukov}. The effects of the initial electron beam position on self-injection of plasma electrons are studied. Employing particle-in-cell (PIC) simulations, it is figured out by injection of a 280 pC bunch in the decelerating phase of 48 TW laser wakefield a high quality (25 fs), high charge (340 pC ), 1 GeV electron bunch is produced.
\begin{figure}
	\centering
	\begin{subfigure}[b]{0.49\linewidth}
		\includegraphics[width=4.2cm,height=3.7cm]{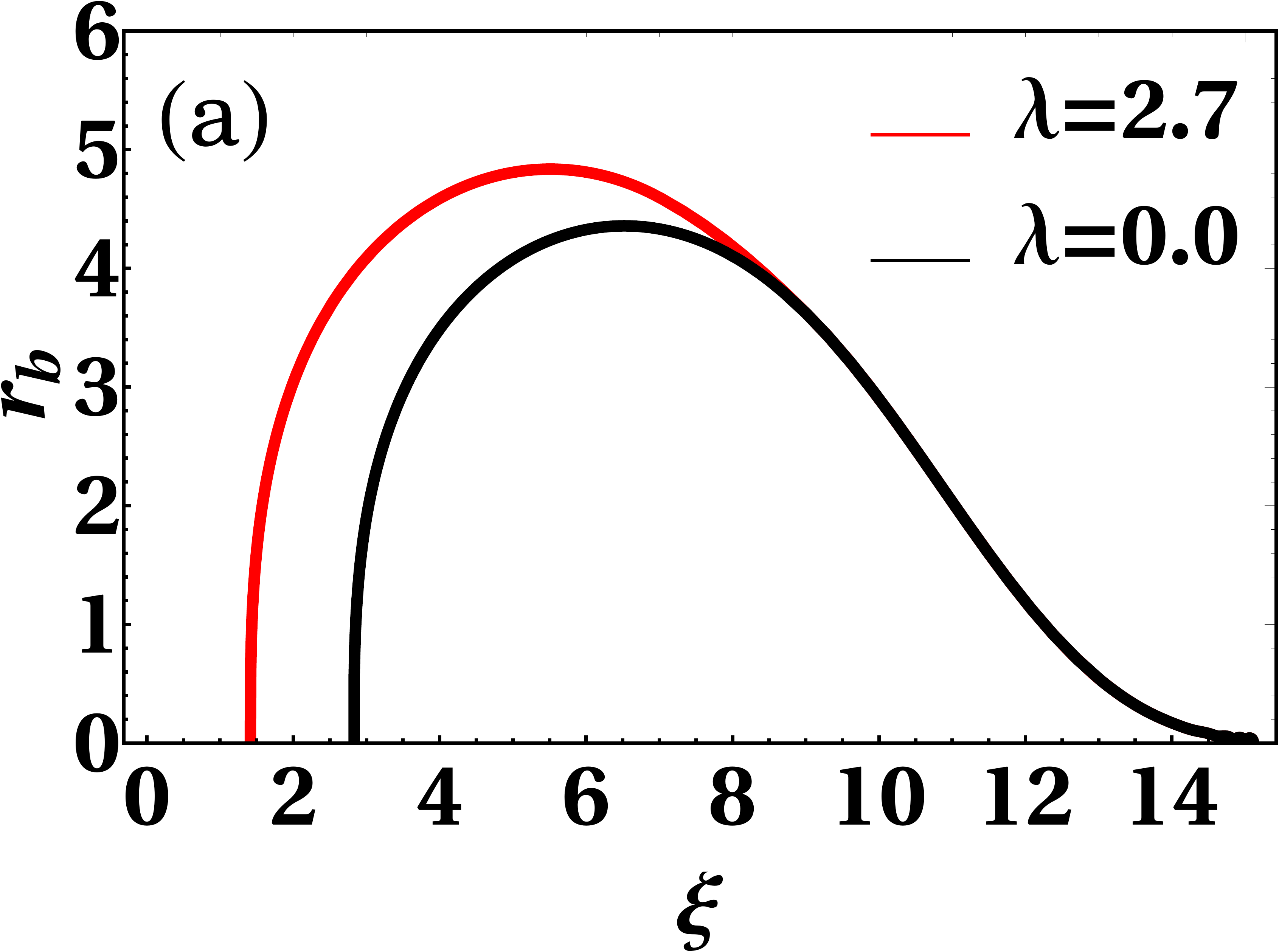}
		
	\end{subfigure}
	\begin{subfigure}[b]{0.49\linewidth}
		\includegraphics[width=4.2cm,height=3.7cm]{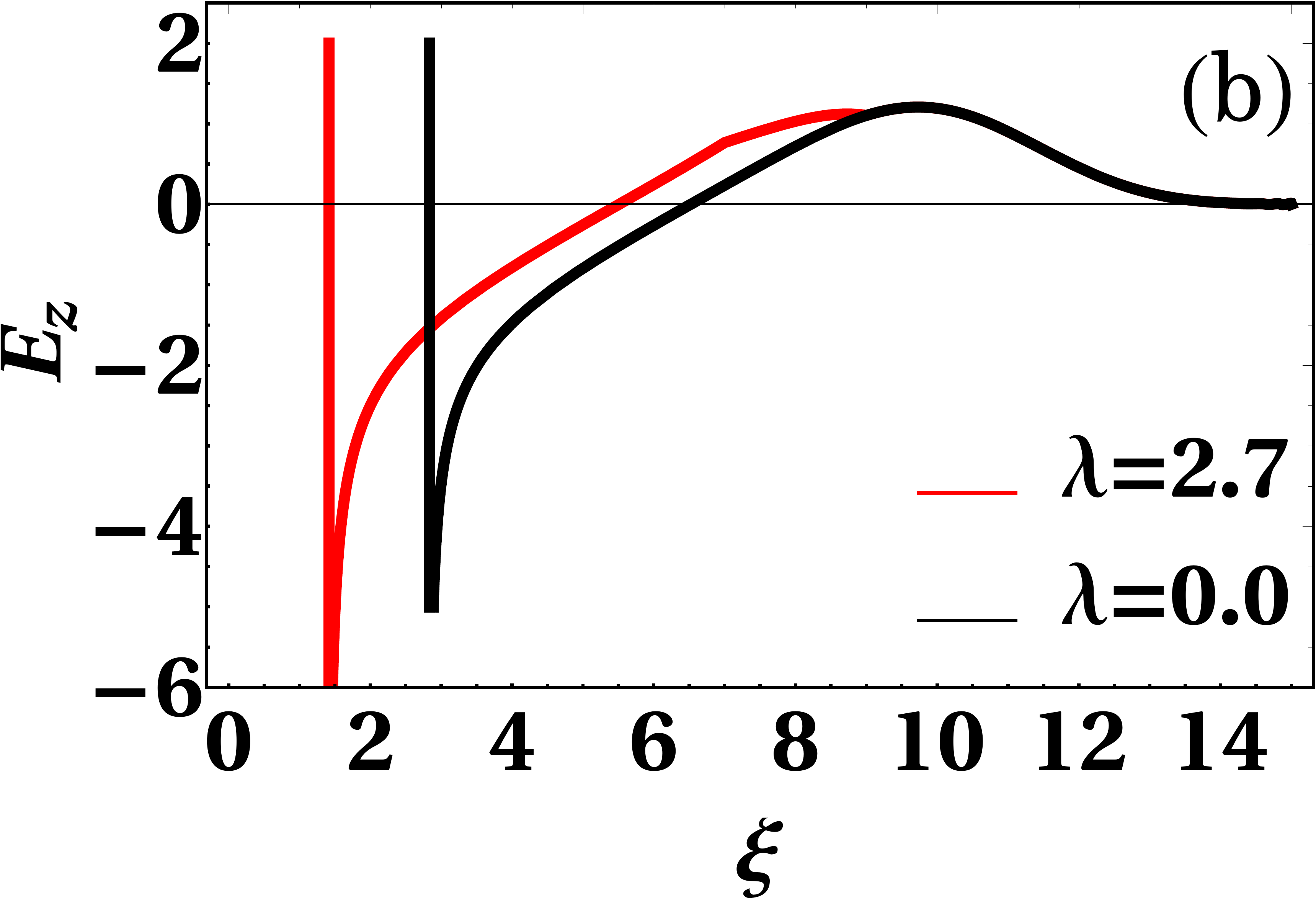}
		
	\end{subfigure}
	\caption{(a) Shape of the bubble and (b) the corresponding wakefields for maximum blowout radius, $r_b = 4$, and a bi-flat-top electron beam with $\lambda = 2.7$. The bunch located at $7\leq\xi\leq9$.}
	\label{Theory}
\end{figure}

We first must represent a way to specify appropriate electron bunch parameters in the hybrid scheme which lead to optimum condition to maximize witness electron energy gain. In fact, the applicability of injecting external charge in the decelerating phase of LWFAs is limited due to transition to the beam-dominated regime. In this regime electrons trajectories in bubble sheath and, in consequence, the wakefield strength dominantly determined by the electron beam.  

We first start with the quantitative analysis of bubble which has an electron bunch in its decelerating phase. It is possible to characterize bubble shape by the trajectory of innermost electrons sheath surrounding the near-spherical cavity \cite{lu1,tzoufras}. The differential equation which describes this structure is studied extensively in Ref. \cite{tzoufras}. In the ultra-relativistic regime, the shape of the sheath is described by 
\begin{equation}\label{1}
r_{b}\frac{d^{2}r_{b}}{d\xi^{2}}+2(\frac{dr_{b}}{d\xi})^{2}+1=4\frac{\lambda(\xi)}{r_{b}^{2}},
\end{equation}
where $r_{b}(\xi)$ defines normalized bubble radius, $\xi$ is the dimensionless coordinate in a frame moving with light velocity and $\lambda(\xi)$ represents the normalized charge per unit length of the load. By direct integrating of Eq. (\ref{1}) some aspects of the evolution of the excited plasma wakefield that happens due to loading electron bunch in the decelerating portion of the plasma bubble can be analyzed. 

Figure \ref{Theory} shows the synergistic nature of the hybrid laser and electron beam drivers.  It is apparent from Fig. \ref{Theory}(a) how the bubble structure and its accelerating field, Fig. \ref{Theory}(b), are changed after loading a bi-flat-top beam within the front part of the  bubble. The presence of electron bunch in the decelerating phase of the wakefield repels the electrons in the sheath around the bubble and makes it bends away. As a result, bubble absorbs the energy of the beam and its maximum radius and the resulted accelerating field grow up. Witness electrons, traveling at nearly c,
move forward in the wake before it phase slip by one-half period of the plasma wave. Therefore, the average field that an electron feels is half of the  useful accelerating field peak in a hybrid laser and bunch wakefield accelerator \cite{lu3}. On the other side, the useful acceleration field an electron feels in the wake of a low-charged relativistic electron beam is \cite{joshi} $E_{z-max}\approx (236\frac{MV}{m})(\frac{N}{4\times10^{4}}){(\frac{600 \mu m}{\sigma_z})}^2\ln (\frac{50 \mu m	}{\sigma  _r}\sqrt{\frac{10^{16} cm^{-3}}{n_0}})$,
where, $n_0$ is plasma density, N, $\sigma_r$ and $\sigma_z $ are number of electrons, transverse spot size and electron beam length, respectively.
Hence, using this analysis, the electron and laser driver parameters can be determined in a way that the average field which an electron feels in a hybrid accelerator is greater than the maximum PWFAs accelerating field for the same acceleration length.

\begin{figure}
	\centering
	\begin{subfigure}[b]{0.49\linewidth}
		\includegraphics[width=4.6cm,height=4.7cm]{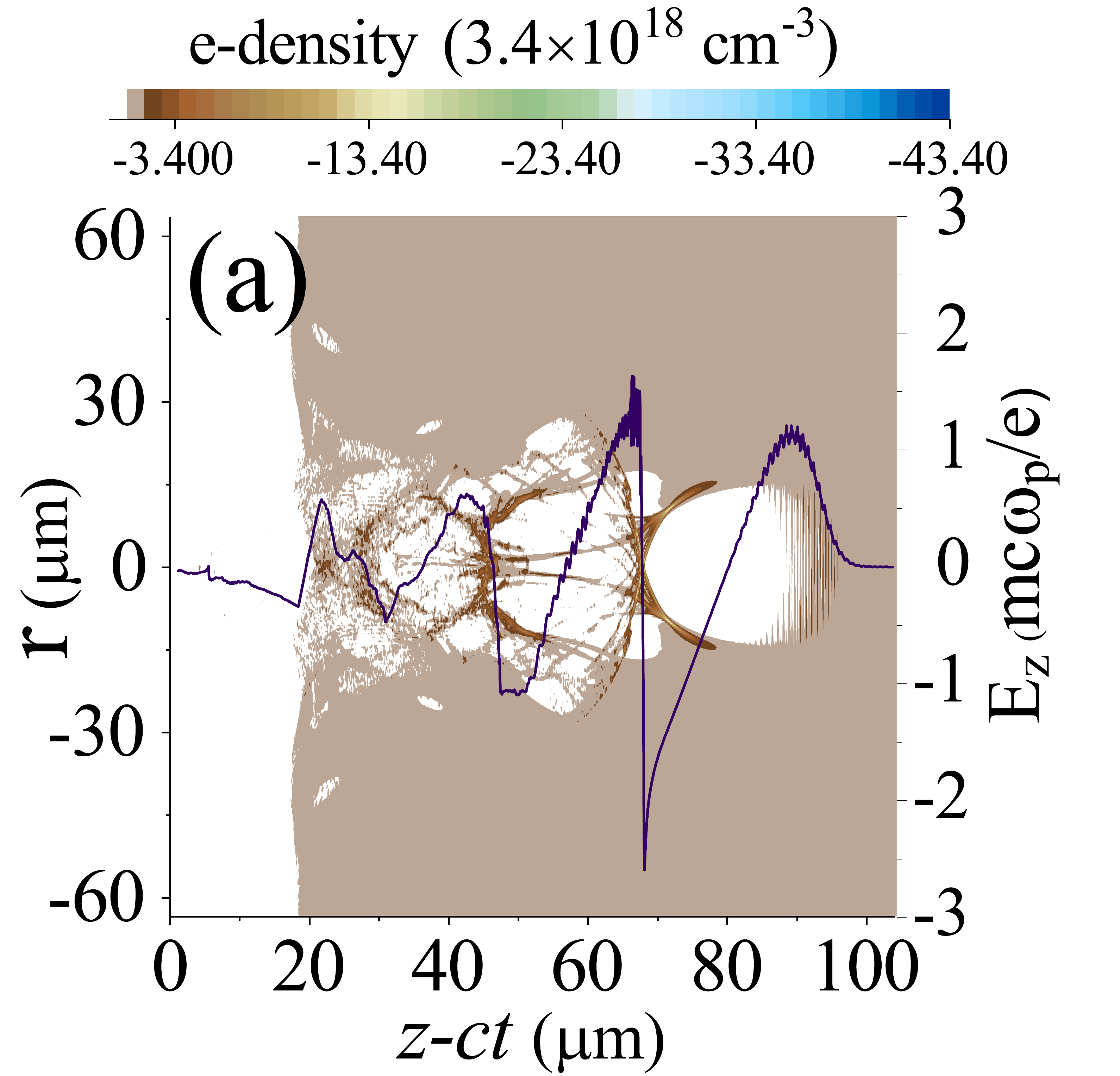}
		
	\end{subfigure}
	\begin{subfigure}[b]{0.49\linewidth}
		\includegraphics[width=4.5cm,height=4.7cm]{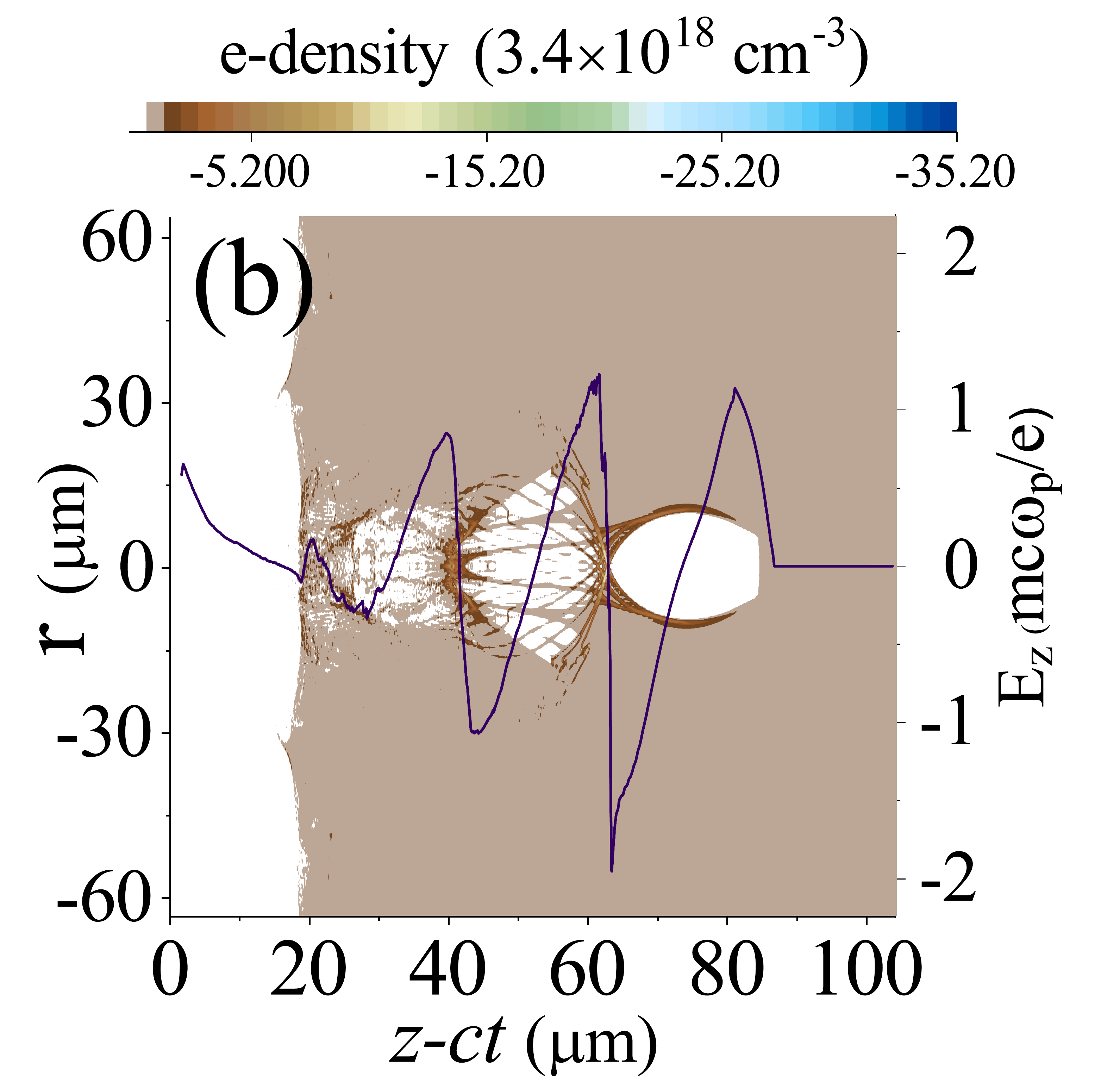}
		
	\end{subfigure}
	\begin{subfigure}[b]{0.49\linewidth}
		\includegraphics[width=4.6cm,height=4.7cm]{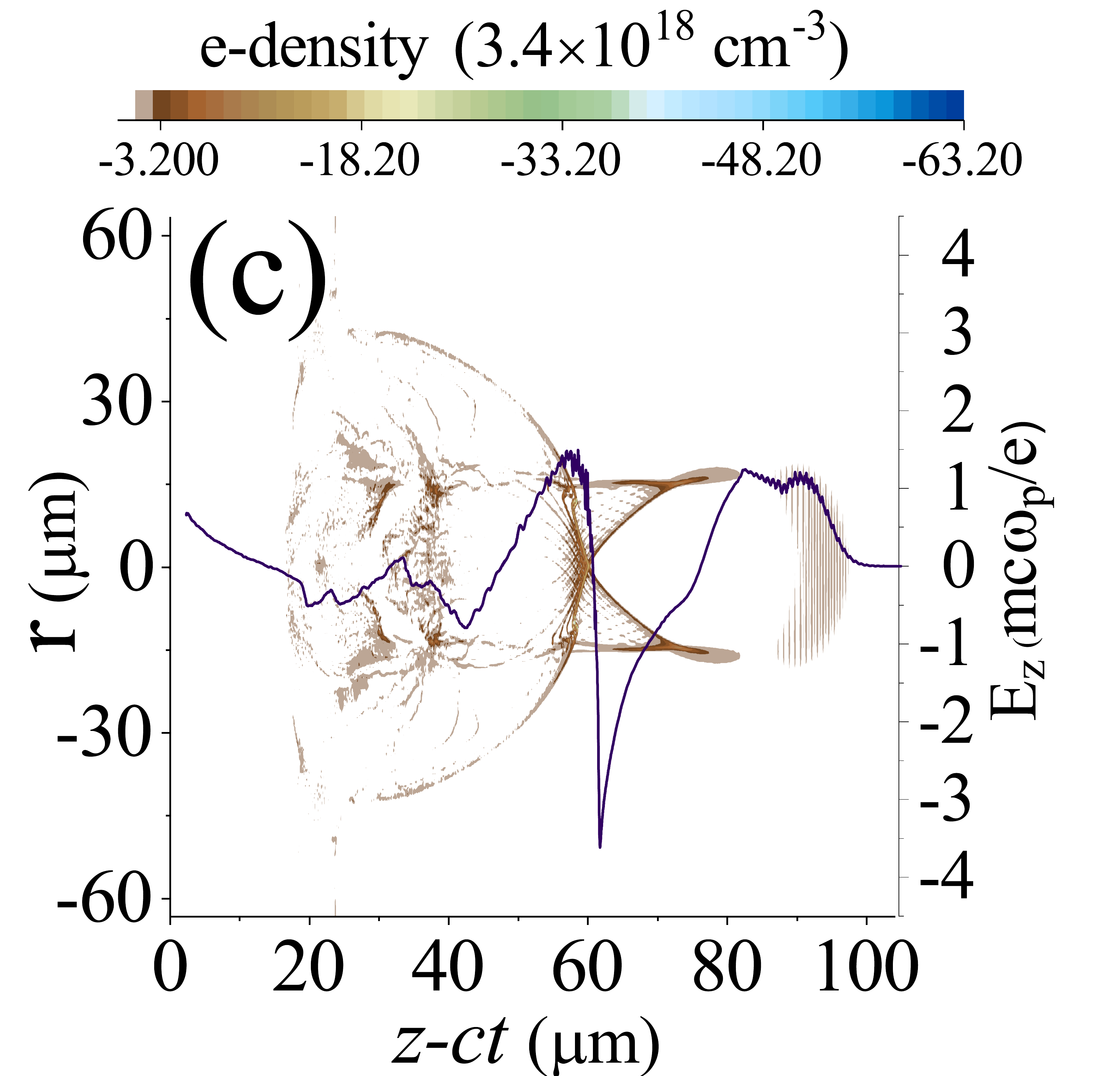}
		
	\end{subfigure}
	\begin{subfigure}[b]{0.49\linewidth}
		\includegraphics[width=4.5cm,height=4.8cm]{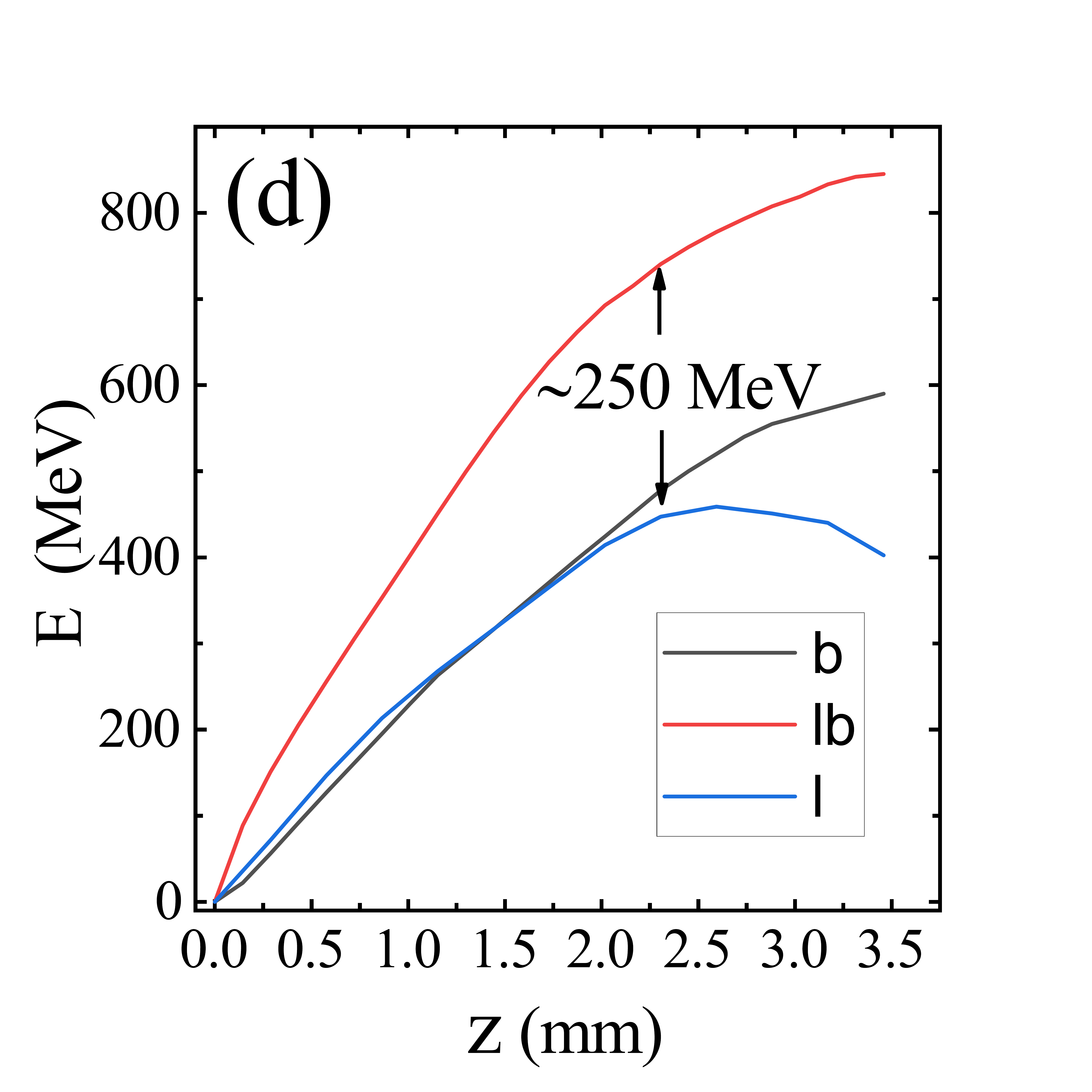}
		
	\end{subfigure}
	\caption{ Background density snapshots and corresponding wakefield after 0.1 mm propagation distance for (a) LWFAs, (b) PWFAs and (c) hybrid laser and electron beam accelerator. (d) Energy gain from the wake of laser driver (blue), electron driver (black), and hybrid driver (red).}
	\label{Theory1}
\end{figure}

To study hybrid laser and electron beam accelerator and check the consequence of injecting an electron bunch into the decelerating phase of LWFAs, we perform a series of 2D simulations using particle-in-cell (PIC) code OSIRIS \cite{osiris}. We consider the bubble regime of electron acceleration. A 30 fs (FWHM), 0.8 $\mu m$, 46 TW laser beam is focused to a Gaussian diffraction-limited spot size of $w_0$ = 52 $\mu$ m which corresponds to normalized vector potential of $a_0 = 4$ at the entrance of plasma density $n_0=3.4\times10^{-18}$ $cm^{-3}$. The parameters of the electron beam are: $k_p\sigma_z$= 2.0, $k_p\sigma_r= $1.5 and $n_{0b}/n_{0} $= 1.2, where $n_{0b}$ is electron beam peak density and $k_p$ is plasma wave number. The electron total charge is estimated at about 0.35 nC and it is behind the peak of the pulse . Initial electron energy is 700 MeV. The simulation window moves at light velocity (c) along the laser propagation direction and have dimensions of 104 $\mu m$ $\times$ 126.7$\mu m$. The number of grid points is 6500 $\times$ 350. Four macro-particles per cell is considered.

To compare the synergistic nature of laser pulse and electron beam  in wakefield accelerators with LWFAs and PWFAs, we displayed snapshots of the background plasma density and corresponding wake electric field  at propagation distance  of 0.1  mm for LWFA, Fig. \ref{Theory1}(a), PWFA, Fig. \ref{Theory1}(b), and hybrid laser and electron beam driver, Fig. \ref{Theory1}(c). It is obvious, how the initial injection in the decelerating phase makes the whole bubble becomes larger.  As a result of whole bubble expansion, stronger charge separation occurs. So, the corresponding accelerating field increases substantially. The co-action of laser and electron beam produces a peak gradient of 530 GeV/m which is about 180 GeV/m larger than each laser or beam driver accelerating field. The transformer ratio (R) which is defined as the maximum accelerating field to the maximum decelerating field increases to R = 3.2  for a hybrid accelerator which is enhanced by a factor of 1.5 and 1.8 compare with LWFAs and PWFAs, respectively. Energy gain of test electrons with initial energy 700 Mev in these three regimes are compared in Fig. \ref{Theory1}(d) for 3.5 mm propagation distance. As it is shown, electrons gain more energy, approximately 250 MeV, in a hybrid accelerator. Practically, acceleration length is limited  to 2.3 mm because electrons in a beam driver lose their energy by the wakefield and start slipping backward and diffracting depleted laser pulse can not be guide stable any more Fig. \ref{Theory1}(d).  

The mechanism of using the hybrid laser pulse and beam driver introduces a new scheme for plasma electron self-injection. In the rest of this letter, we will investigate this new self-injection scenario. The first electron bunch slips forward with respect to the laser so the bubble radially and longitudinally expands. This extension reduces the gamma-factor of the bubble. In addition, the wakefield gets stronger as a result of beam and laser synergy, so the trapping of electrons at lower plasma densities and laser powers becomes attainable. By specifying the appropriate range of parameters and initial injection position for electron bunch driver, producing a high amount of charge and low energy spread electron trapping is possible.

\begin{figure}
	\centering       
	\begin{subfigure}[b]{0.49\linewidth}
		\includegraphics[width=5cm,height=5cm]{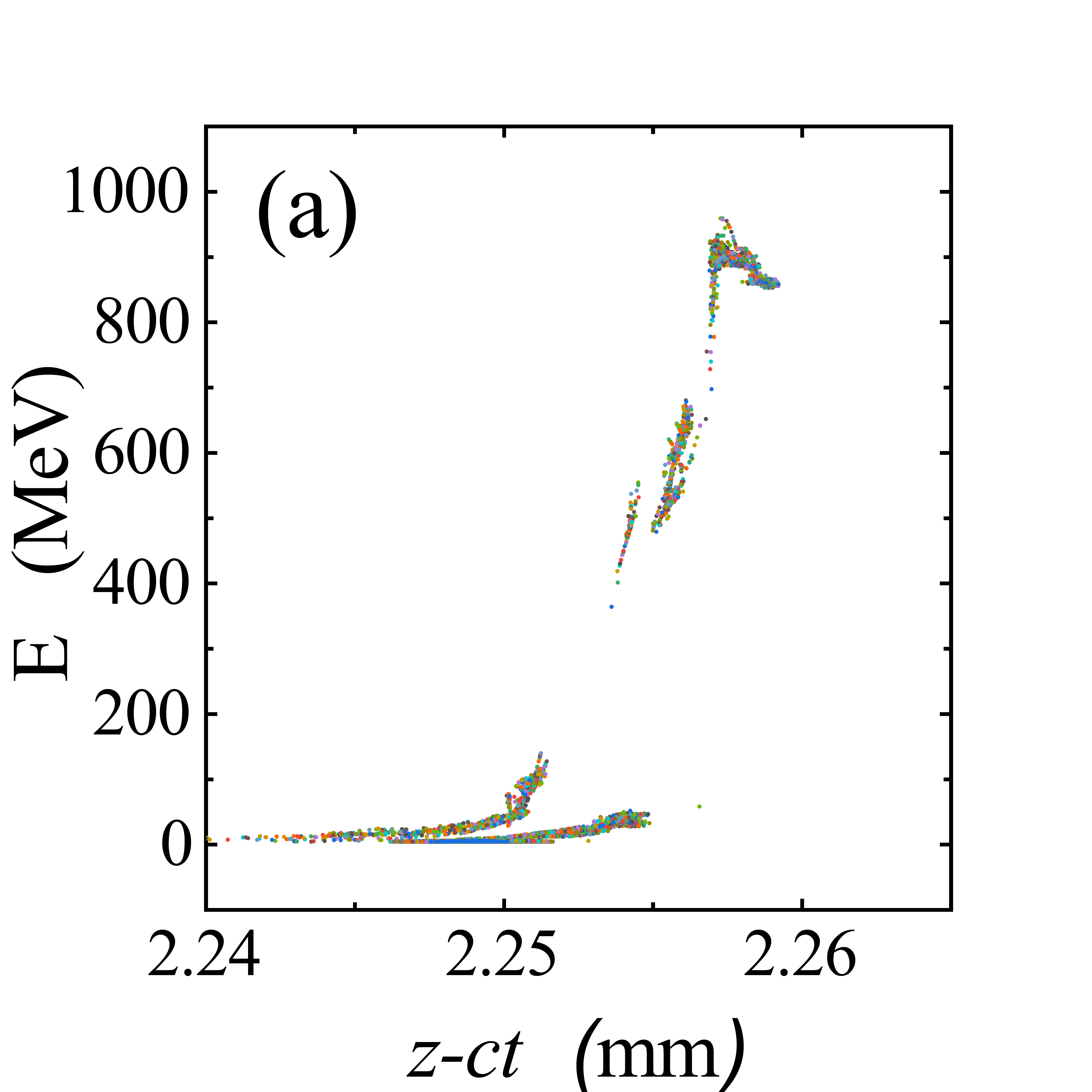}
		
	\end{subfigure}
	\begin{subfigure}[b]{0.49\linewidth}
		\includegraphics[width=5cm,height=5cm]{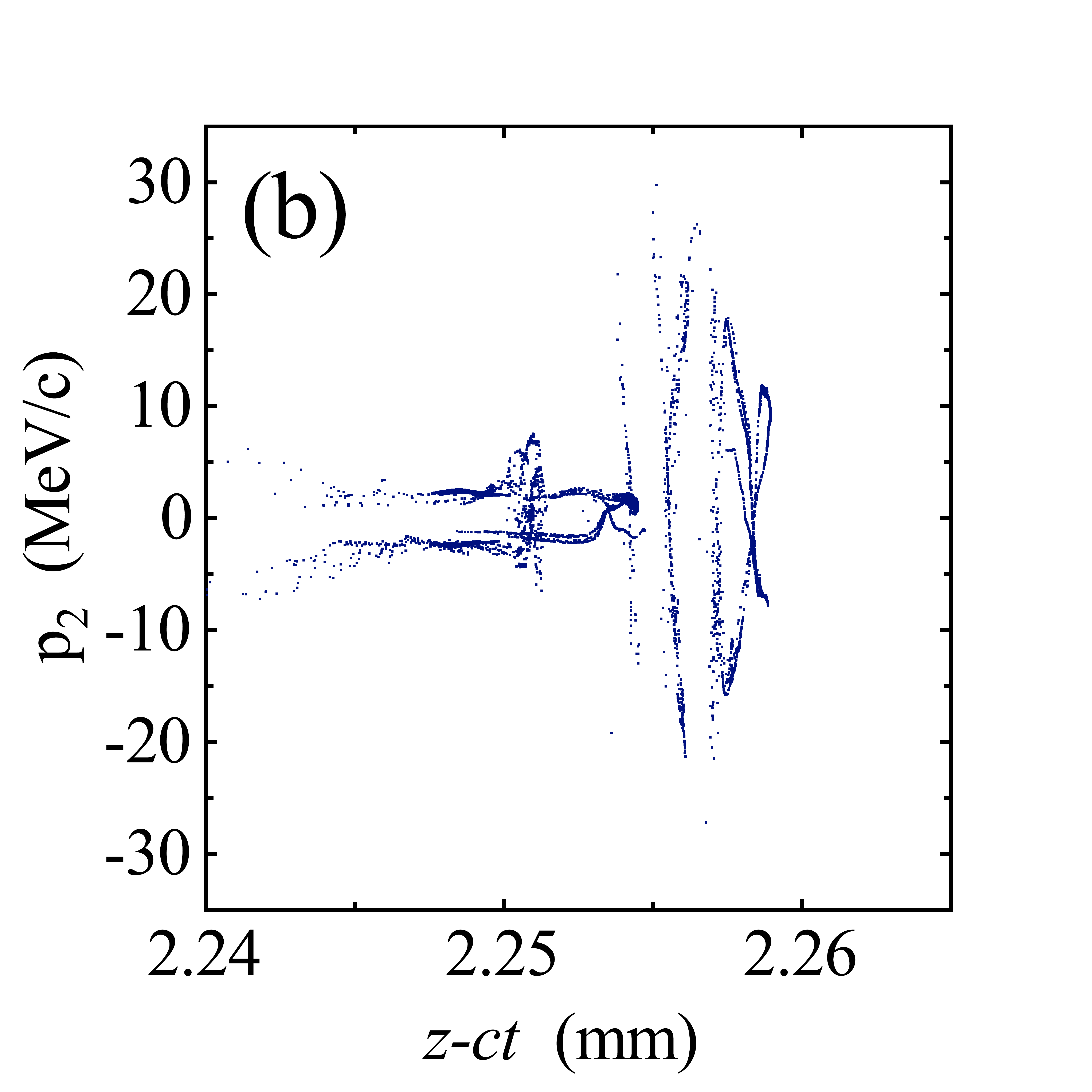}
		
	\end{subfigure}
	\begin{subfigure}[b]{0.49\linewidth}
		\includegraphics[width=5.cm,height=5.cm]{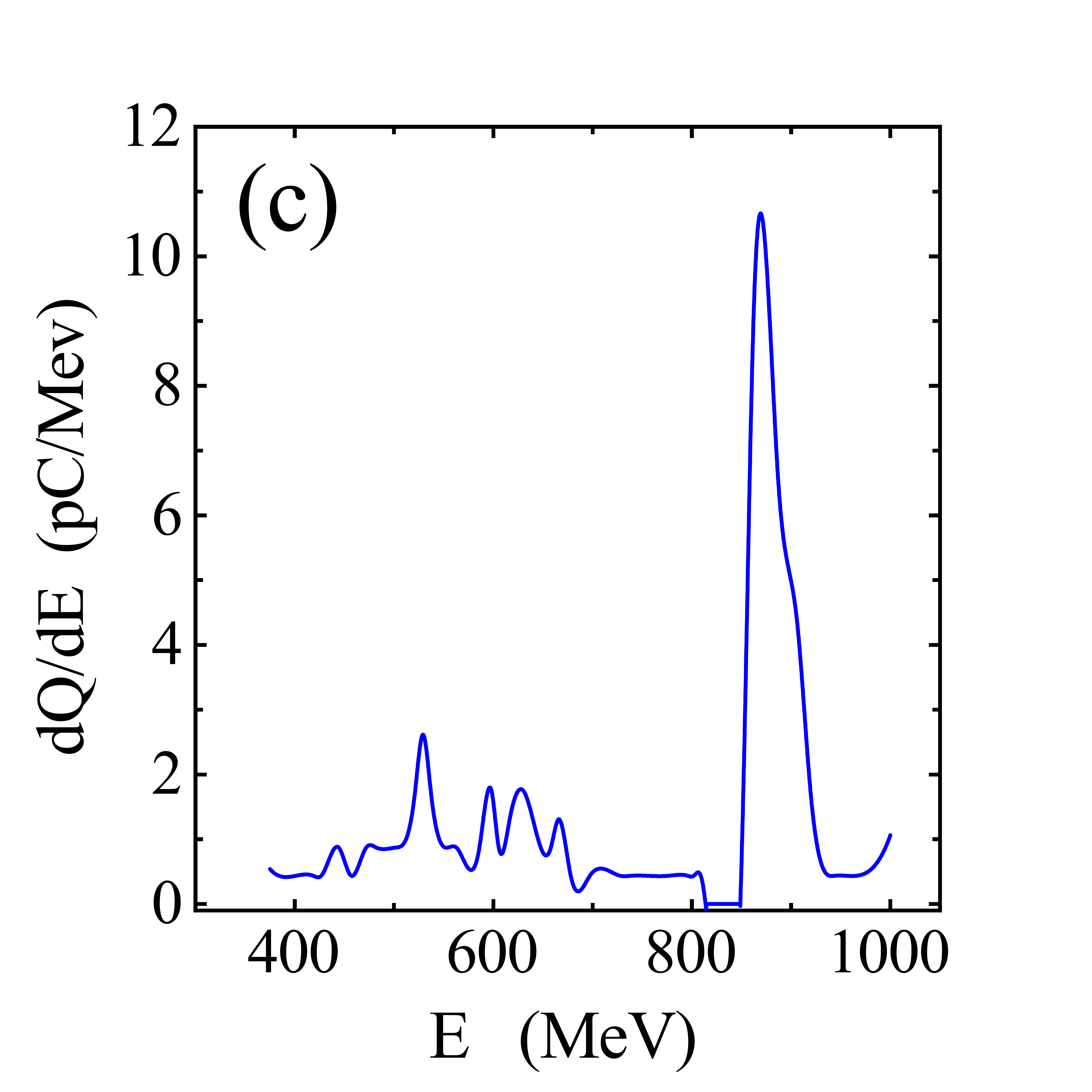}
		
	\end{subfigure}	
	\caption{The phase space of self-trapped electrons for a hybrid accelerator after 2.4 mm  propagation. Energy gain from the wake (a), Transverse momentum vs propagation distance (b), and electron energy spectrum (c).}
	\label{Theory2}
\end{figure}

To illustrate the quality, charge, and energy of self-injected electrons, we plotted phase space and energy spectrum of them in Fig. \ref{Theory2}  at propagation distance 2.3 mm for the same simulation parameters. Figure \ref{Theory2} represents two kinds of self-trapped electrons: a peak with 340 pC of charge and an electron tail that carries 60 pC of charge. Thus, the electron peak forms the principal part of the self-injected electrons. The self-trapped electrons are produced by transverse injection mechanism \cite{corde}. They mostly originate from electron sheath around the bubble. As Fig. \ref{Theory2}(b) confirms the peak electrons have larger transverse momentum while electrons in the tail experience smaller transverse motion. The main peak of accelerated electrons reaches 900 MeV energy with rms spread of 6.6$\%$  and a duration of 25 fs. Therefore, combining an electron bunch of 340 pC with a laser pulse generates an electron beam with 140 times larger amounts of charge. 

\begin{figure}
	\centering
	\begin{subfigure}[b]{0.49\linewidth}
		\includegraphics[width=4.45cm,height=4.5cm]{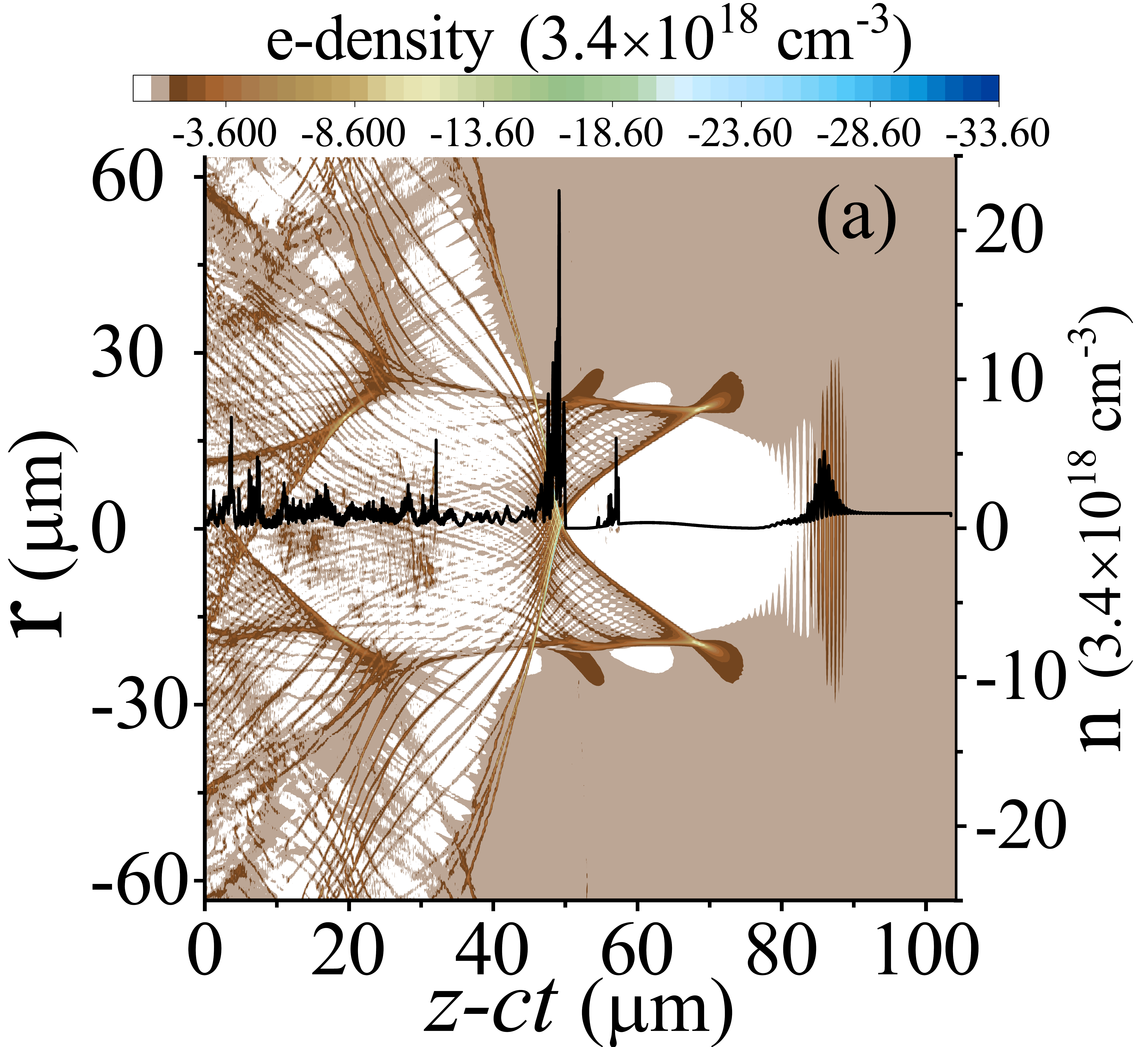}
		
	\end{subfigure}
	\begin{subfigure}[b]{0.49\linewidth}
		\includegraphics[width=4.45cm,height=4.5cm]{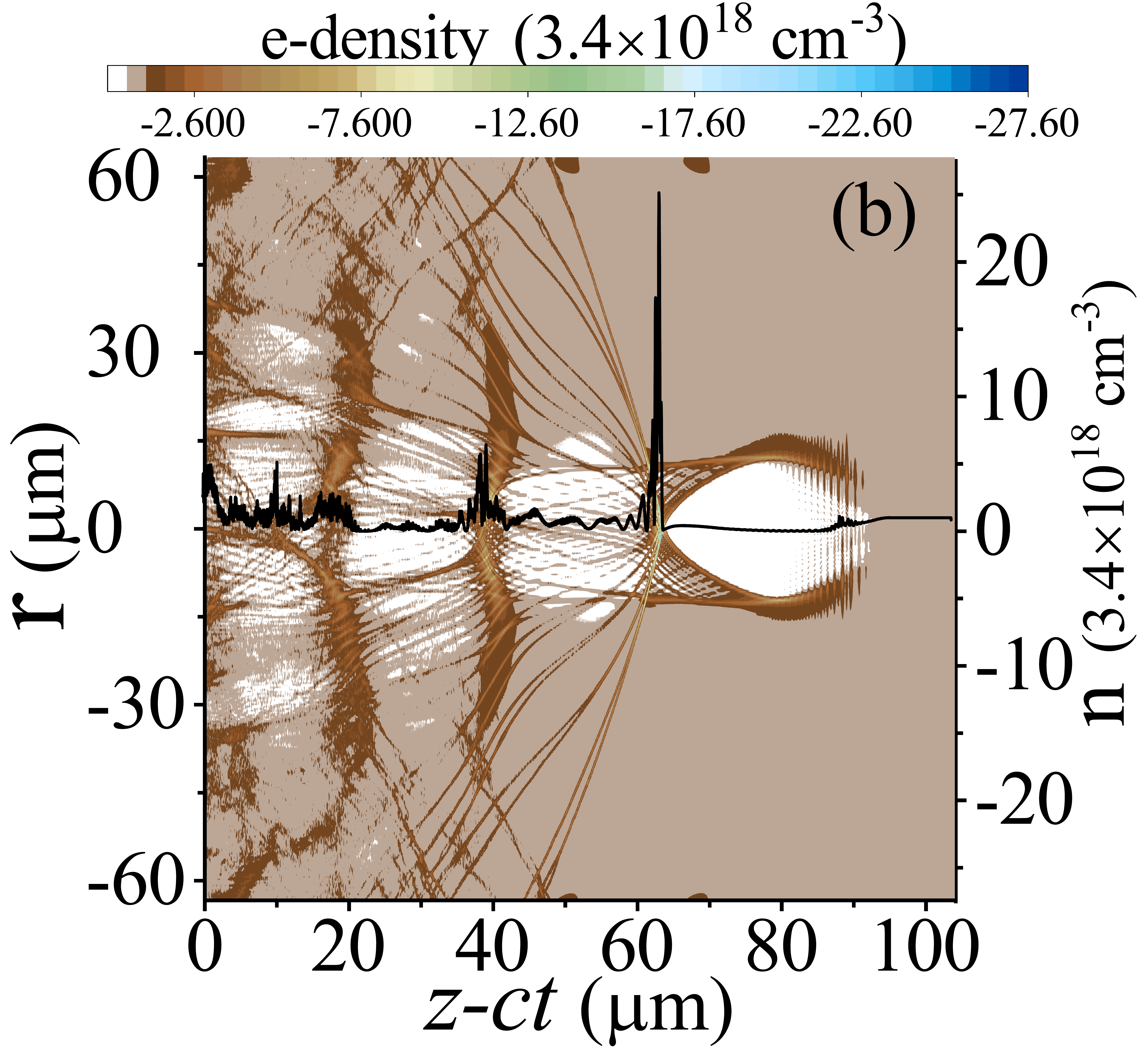}
		
	\end{subfigure}
	\begin{subfigure}[b]{0.49\linewidth}
		\includegraphics[width=4.6cm,height=4.5cm]{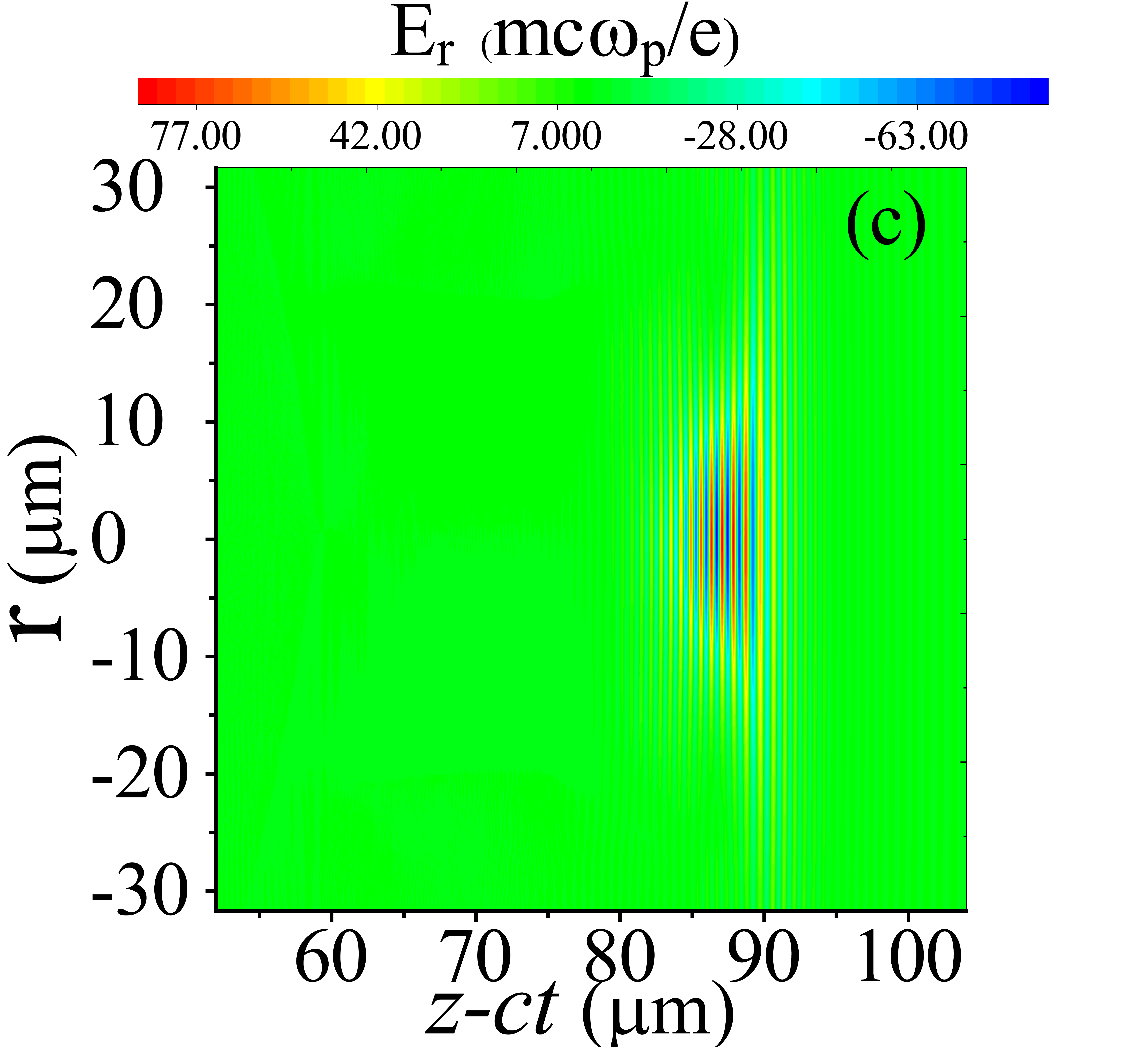}
		
	\end{subfigure}
	\begin{subfigure}[b]{0.49\linewidth}
		\includegraphics[width=4.5cm,height=4.5cm]{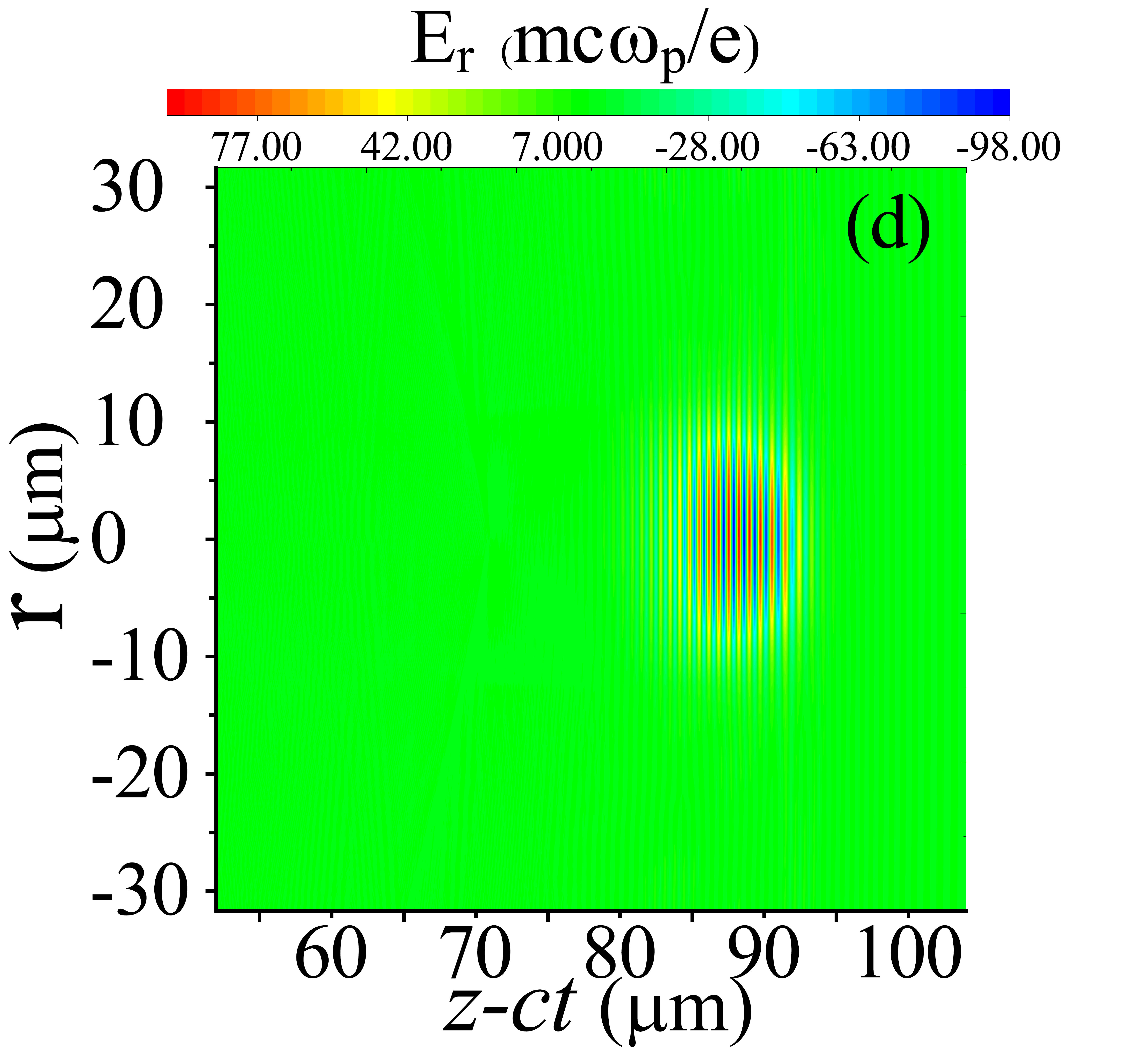}
		
	\end{subfigure}
	\caption{Effects of initial electron beam position on electron self-injection and laser self-guiding at 2.3 mm propagation distance. (a) Plasma electron density snapshot and density distribution on propagation axis (black line). The electron beam is located behind the laser pulse peak. (b) Plasma electron density snapshot and density distribution on the propagation axis (black line). The electron beam is located in front of the laser pulse peak. (c) Transverse laser pulse profile. Electron beam is located behind the laser pulse peak. (d) Transverse laser pulse profile. The electron beam is located in front of the laser pulse peak.}
	\label{Theory3}
\end{figure}

The initial beam  position makes a major contribution in  self-trapping of electrons in the bubble wakefield. Locating the electron beam  behind the peak of laser pulse gives rise to self-trapping of electrons because the presence of an electron beam in the decelerating part of the wakefield makes an evolving bubble as discussed before, Fig. \ref{Theory3}(a). However, if the electron beam is located in the front part of the laser pulse, it contributes more to repelling plasma electrons outward and forming the wider plasma ion channel behind, Fig. \ref{Theory3}(b). Figure \ref{Theory3}(d) confirms that when the electron beam is located in the front part of the pulse, the laser stayed more focused and can be self-guided for more propagation distances. Although, the laser pulse is almost diffracted in the same propagation distance when the electron beam is behind the peak of the laser pulse, Fig. \ref{Theory3}(c).

In conclusion, a new type of wakefield  accelerator using laser pulse and electron beam synergy is presented. An interpretive model is introduced to determine applicable laser and beam parameters. In this scheme, the injection of 280 pC electron beam charge in decelerating part of wakefield causes the bubble to expands, which then strengthens the accelerating field and gives rise to self-injection at lower plasma densities and laser powers. The witness beam gained 250 MeV more energy in a hybrid accelerator compare with LWFAs and PWFAs. The self-trapped electrons are very short and accelerated to near-Gev range with low energy spread and high charge. Studding the initial electron beam position demonstrates that electron self-trapping occurs if the beam is located behind the peak of the laser pulse.

The authors would like to acknowledge the OSIRIS Consortium, consisting of UCLA and IST (Lisbon, Portugal) for providing access to the OSIRIS 4.0 framework.

\bibliography{s.bib}

\end{document}